\documentclass[12pt]{article}

\usepackage{amsmath,amssymb,amsfonts,amsbsy}
\usepackage{cite}
\usepackage{graphicx}
\usepackage{epsfig}

\usepackage{bm}
\usepackage{color}                                                        

\DeclareFontFamily{OT1}{mygreek}{}%
\DeclareFontShape{OT1}{mygreek}{m}{n}{<->omsegr}{}%
\DeclareFontShape{OT1}{mygreek}{b}{n}{<->omsegrb}{}%
\DeclareFontShape{OT1}{mygreek}{m}{it}{<->omsegri}{}%
\DeclareFontShape{OT1}{mygreek}{bx}{n}{<->sub * mygreek/b/n}{}%
\DeclareFontShape{OT1}{mygreek}{m}{sl}{<->sub * mygreek/m/it}{}%
\DeclareSymbolFont{Greekrm}{OT1}{mygreek}{m}{n} 
\DeclareSymbolFont{Greekbf}{OT1}{mygreek}{b}{n} 
\DeclareSymbolFont{Greekit}{OT1}{mygreek}{m}{it} 
\DeclareMathSymbol{\omegab}{\mathalpha}{Greekbf}{119}

\begin{document}
\addcontentsline{toc}{subsection}{{Spinning particles in de Sitter spacetime}\\
{\it Yu.N. Obukhov}} 

\setcounter{section}{0}
\setcounter{subsection}{0}
\setcounter{equation}{0}
\setcounter{figure}{0}
\setcounter{footnote}{0}
\setcounter{table}{0}

\begin{center}
\textbf{SPINNING PARTICLES IN DE SITTER SPACETIME}

\vspace{5mm}

\underline{Yu.N.~Obukhov}$^{\,1\,\dag}$ and D.~Puetzfeld$^{\,2\,\ddag}$

\vspace{5mm}

\begin{small}
  (1) \emph{Inst. Theoretical Physics, University of Cologne, Z\"ulpicher Str. 77, 50937 K\"oln, Germany}\\
  (2) \emph{ZARM, University of Bremen, Am Fallturm, 28359 Bremen, Germany} \\
  $\dag$ \emph{E-mail: yo@thp.uni-koeln.de}
  $\ddag$ \emph{E-mail: dirk.puetzfeld@zarm.uni-bremen.de}
\end{small}
\end{center}

\vspace{0.0mm} 

\begin{abstract}
We report on the multipolar equations of motion for spinning test bodies in the de Sitter spacetime of constant positive curvature. The dynamics of spinning particles is discussed for the two supplementary conditions of Frenkel and Tulczyjew. Furthermore, the 4-momentum and the spin are explicitly expressed in terms of the spacetime coordinates with the help of the 10 Killing vectors available in de Sitter spacetime.
\end{abstract}

\vspace{7.2mm} 

The multipolar equations of motion, commonly termed Mathisson-Papapetrou equations \cite{DPYNO_Mathisson:1937,DPYNO_Papapetrou:1951:3,DPYNO_Tulczyjew:1959,DPYNO_Tulczyjew:1962,DPYNO_Dixon:1964,DPYNO_Corben:1968,DPYNO_Madore:1969,DPYNO_Dixon:1974:1,DPYNO_Dixon:1979,DPYNO_Puetzfeld:Obukhov:2007,DPYNO_Dixon:2008:1}
\begin{eqnarray}
\dot{p}^\alpha = -\,{\frac 12}\,S^{\mu\nu}u^\beta\,R_{\mu\nu\beta}{}^\alpha, \quad \quad \dot{S}^{\alpha\beta}  =  2p^{[\alpha}\,u^{\beta]}, \label{mpd_equations}
\end{eqnarray}
represent a self-consistent set of equations, which is widely used for description of spinning test bodies in General Relativity. Here $``\dot{\phantom{aa}}"= D/ds = u^\alpha\nabla_\alpha$, and a test particle is described in terms of the 4-momentum $p^\alpha$, the 4-velocity $u^\alpha$, and the tensor of spin $S^{\alpha \beta}$. We report on the solution of the equations of motion (\ref{mpd_equations}) in a maximally symmetric 4-dimensional space represented by de Sitter spacetime. Our analysis extends well-known results from flat spacetime -- in which the dynamics of spinning test bodies already becomes nontrivial \cite{DPYNO_Frenkel:1926,DPYNO_Weyssenhoff:Raabe:1947,DPYNO_Kudryashova:Obukhov:2010} -- to a manifold with non-vanishing curvature. In particular we make use of the two frequently used supplementary conditions \cite{DPYNO_Frenkel:1926,DPYNO_Pirani:1956,DPYNO_Tulczyjew:1959}
\begin{equation}
S^{\alpha\beta}u_\beta = 0, \quad (\ast), \quad \quad S^{\alpha\beta}p_\beta = 0. \quad (\ast\ast) \label{supp_conditions}
\end{equation}
and are able to obtain analytic results. This is in contrast to the analysis in most other spacetimes in which the lack of symmetry -- compared to de Sitter space -- complicates the situation and usually necessitates to make additional simplifying assumptions. We should stress that our analysis is valid for both interpretations of the Mathisson-Papapetrou equations which can be found in the literature, i.e.\ it applies to point particles as well as to extended test bodies. One should keep in mind though, that there are preferences regarding the supplementary condition \cite{DPYNO_Beiglboeck:1965,DPYNO_Beiglboeck:1967,DPYNO_Schattner:1978,DPYNO_Schattner:1979}, depending on the system which is supposed to be described by the equations of motion. In the following our conventions and notation follows that of \cite{DPYNO_Puetzfeld:Obukhov:2011}.

\paragraph{Spacetime with maximal symmetry} The curvature of de Sitter spacetime is given by $R_{\mu\nu\alpha}{}^\beta = \frac{1}{\ell^2} \left(g_{\alpha\mu}\delta^\beta_\nu - g_{\alpha\nu}\delta^\beta_\mu\right)$, where $\ell$ is a real constant. The first equation of motion from (\ref{mpd_equations}) then reduces to 
\begin{equation}
\dot{p}^\alpha = {\frac 1 {\ell^2}}\,S^{\alpha\beta}u_\beta.\label{dPS}
\end{equation}

\paragraph{Tulczyjew condition}
Assuming ($\ast\ast$) from (\ref{supp_conditions}), we introduce the 4-vector of spin via $\check{S}^\alpha := \eta^{\alpha\beta\mu\nu}p_\beta S_{\mu\nu}$. The inverse formula yields the spin tensor in terms of the spin vector: $S^{\alpha\beta} ={\frac {1}{2M^2}} \eta^{\alpha\beta\mu\nu}p_\mu\check{S}_\nu$, with $M^2:=p^\alpha p_\alpha$. As a result, the equations of motion of a spinning test body in the de Sitter spacetime under the Tulczyjew condition reduce to
\begin{equation}
\dot{p}^\alpha \stackrel{(\ast\ast)}{=} 0,\qquad \dot{\check{S}}{}^\alpha \stackrel{(\ast\ast)}{=} 0,\qquad p^\alpha \stackrel{(\ast\ast)}{=} m\,u^\alpha, \label{em1}
\end{equation}
where we introduced $m:=p^\alpha u_\alpha$, or, equivalently
\begin{equation}
\dot{u}^\alpha \stackrel{(\ast\ast)}{=} 0, \qquad \eta^{\alpha \beta \gamma \delta} u_\beta \dot{S}_{\gamma \delta} \stackrel{(\ast\ast)}{=} 0.
\end{equation}
The first equation actually means that the trajectories of the spinning bodies are the geodesics in the de Sitter space. The second equation describes the precession of the spin vector, or tensor, of a body during its motion along a geodesic curve. 

\paragraph{Frenkel condition}
The dynamics for ($\ast$) from (\ref{supp_conditions}) have a certain similarity to the above case, however there are important differences. In particular, from (\ref{dPS}) we immediately infer that, like in the previous case, the momentum is covariantly constant, $\dot{p}^\alpha = 0$. Following the same line of reasoning, we define the 4-vector of spin by ${S}^\alpha := \eta^{\alpha\beta\mu\nu}u_\beta S_{\mu\nu}$. The inverse formula yields the spin tensor in terms of the spin vector: $S^{\alpha\beta} = {\frac{1}{2}}\eta^{\alpha\beta\mu\nu}u_\mu{S}_\nu$ (we use the normalization $u^2 = 1$). Directly from the definition of the spin vector, we derive that the spin vector is Fermi-Walker transported. We thus have the system
\begin{equation}\label{em2}
\dot{p}^\alpha = 0,\qquad \rho^\alpha_\beta \dot{S}^\beta = 0.
\end{equation}
Here $\rho^\alpha_\beta:=\delta^\alpha_\beta - u^\alpha u_\beta$.
Although this looks formally similar to (\ref{em1}), the actual dynamics is very different. In particular, the trajectories are no longer geodesics because the momentum does not coincide with the velocity. The above system can be simplified even further and we end up with the final system:
\begin{eqnarray}
\dot{p}^\alpha \stackrel{(\ast)}{=} 0,\quad \dot{S}^\alpha \stackrel{(\ast)}{=} 0,\quad p^\alpha \stackrel{(\ast)}{=} mu^\alpha - {S}^{\alpha\beta} \dot{u}_\beta. \label{em_frenkel_final}
\end{eqnarray}
Hence, in de Sitter space the spin is also parallely transported under the Frenkel condition. Equivalently, one may look for solutions of the system
\begin{eqnarray}
S^{\alpha \beta} \ddot{u}_\beta - m \dot{u}^\alpha \stackrel{(\ast)}{=} 0,\quad \quad \dot{S}^{\alpha \beta} + 2 u^{[\alpha} S^{\beta]\gamma} \dot{u}_\gamma \stackrel{(\ast)}{=}0.
\end{eqnarray}
Geodetic motion plus parallel transport of the spin ($\dot{u}^\alpha = 0$, $\dot{S}^\alpha = 0$) is a solution of (\ref{em_frenkel_final}). However, in general the motion of a spinning body described by (\ref{em_frenkel_final}) is more complicated. We introduce a new vector variable $Q^\alpha := {\frac 1{M^2}}\,S^{\alpha\beta}p_\beta$, which has constant length and is orthogonal to the velocity -- i.e.\  $Q^\alpha$ is spacelike. One can then show, that the second derivative of $Q^a$ is actually the force which pushes the body away from the geodesic, i.e.\ in general one has $\dot{u}{}^\alpha = -\,\ddot{Q}{}^\alpha$. Furthermore the ``$Q$-force'' fulfills an oscillator equation 
\begin{equation}
\ddot{Q}{}^\alpha + \omega^2\,Q^\alpha = 0,\label{oscQ}
\end{equation}
with the frequency $\omega := {{2M}/{\sqrt{-S_\alpha S^\alpha}}} = {{M}/{\sqrt{\frac{1}{2}S_{\alpha\beta}S^{\alpha\beta}}}}$. Qualitatively, the dynamics of spinning bodies subject to the Frenkel condition in the de Sitter spacetime is similar to that in flat space \cite{DPYNO_Kudryashova:Obukhov:2010}. Everything is determined by the initial conditions. If initially (at the proper time $s=0$) spin is parallel to the momentum, i.e. $S^{\alpha\beta}p_\beta =0$ (hence $Q^\alpha =0$), then this is true on the whole trajectory that turns out to be geodesic. Otherwise, the trajectory is a geodesic curve, perturbed by the oscillatory motion of $Q^\alpha$ with the frequency $\omega$.

\paragraph{Integrals of motion}

The de Sitter spacetime has exactly the same number of Killing vectors as the total number of the ``gravitational charges'', that is, 10. Hence, we can try to find the momentum $p^\mu$ and the spin $S^{\mu\nu}$ without solving differential equations by just making use of the 10 conservation laws. This task is most straightforwardly treated in the conformally flat representation, i.e.\ 
\begin{equation}\label{confds}
ds^2 = \varphi^2\eta_{\alpha \beta}dx^\alpha dx^\beta,\qquad \eta_{\alpha \beta} = {\rm diag}(+1,-1,-1,-1).
\end{equation}
The conformal factor depends only on the 4-dimensional ``radius'' $\sigma = \eta_{\alpha \beta} x^\alpha x^\beta$, namely,
\begin{equation}
\varphi = {\frac{1}{1 - {\frac{\sigma}{4\ell^2}}}} = {\frac{1}{1 - {\frac{\eta_{\alpha \beta}x^\alpha x^\beta}{4\ell^2}}}}.\label{confi}
\end{equation}
In this representation, the Killing vectors \cite{DPYNO_Bokhari:Quadir:1987} are as follows: 
\begin{eqnarray}\label{xia}
{\underset {(\alpha)}\xi} &=& \left(1+{\frac {\sigma}{4\ell^2}}\right) \partial_\alpha - {\frac {x_\alpha x^\beta}{2\ell^2}}\,\partial_\beta,\\
{\underset {[\alpha\beta]}\xi} &=& x_\alpha\partial_\beta - x_\beta\partial_\alpha.\label{xiab}
\end{eqnarray}
Furthermore, we make use of the fact that the scalar 
\begin{equation}\label{conserved}
2\xi_\alpha p^\alpha + S^{\alpha\beta}\nabla_\alpha\xi_\beta = {\rm const}
\end{equation}
is an integral of motion of the system (\ref{mpd_equations}) for any Killing vector $\xi^\alpha$. By substituting (\ref{xia}) and (\ref{xiab}) into (\ref{conserved}), we have the algebraic system
\begin{eqnarray}\label{Pi}
2{\underset {(\alpha)}\xi}{\!}_\mu p^\mu + S^{\mu\nu}\nabla_\mu {\underset {(\alpha)}\xi}{\!}_\nu &=& 2\Pi_\alpha,\\ \label{Sigma}
2{\underset {[\alpha\beta]}\xi}{\!}_\mu p^\mu + S^{\mu\nu} \nabla_\mu{\underset {[\alpha\beta]}\xi}{\!}_\nu &=& 2\Sigma_{\alpha\beta}.
\end{eqnarray}
Here $\Pi_\alpha$ and $\Sigma_{\alpha\beta}= -\Sigma_{\beta\alpha}$ are the 4 + 6 = 10 constants of motion. This system is solved by 
\begin{eqnarray}
p^\mu &=& {\frac {1}{\ell^2}}\,\eta^{\mu\alpha}\Sigma_{\alpha\beta}x^\beta + \check{\eta}^{\mu\nu}\,\Pi_\nu,\label{Pdes}\\ 
S^{\mu\nu} &=& \hat{\eta}^{\mu\alpha}\hat{\eta}^{\nu\beta}\,\Sigma_{\alpha\beta} + \hat{\eta}^{\mu\alpha}\Pi_\alpha\,x^\nu - \hat{\eta}^{\nu\alpha}\Pi_\alpha\,x^\mu,\label{Sdes}
\end{eqnarray}
where we introduced $\hat{\eta}_{\mu\nu} := \left(1 - {\frac {\sigma}{4\ell^2}}\right) \eta_{\mu\nu} + {\frac {x_\mu x_\nu}{2\ell^2}}$ and $\check{\eta}_{\mu\nu} := \left(1 + {\frac{\sigma}{4\ell^2}} \right) \eta_{\mu\nu} - {\frac{x_\mu x_\nu}{2\ell^2}}$, i.e.\ we are able to express the momentum and the spin as functions of the constants of motion. Remarkably, the dependence on the spacetime coordinates is merely polynomial. 

\paragraph{Summary} Qualitatively, the dynamics of spinning test bodies in de Sitter spacetime is similar to the one obtained in flat spacetime. For the Tulczyjew condition ($\ast\ast$), the body moves along a geodesic curve, whereas the spin vector is parallelly transported along the trajectory. In the Frenkel case ($\ast$), the spin is still parallelly transported, but geodesic motion is just one special solution of the equations of motion. When the initial value of $Q^\alpha$ is nontrivial, then the body is affected by the spin-dependent force, the acceleration $\dot{u}^\alpha$ is nontrivial, and the trajectory oscillates around a geodesic with the frequency $\omega$. Furthermore, the high symmetry of de Sitter spacetime allows for polynomial expressions -- w.r.t.\ the spacetime coordinates -- of the momentum and spin as functions of the constants of motion.

\paragraph{Acknowledgments}

The authors are grateful to F.W.\ Hehl (Univ.\ Cologne) for stimulating discussions and constructive criticism. The work of Y.N.O. was partly supported by the German-Israeli Foundation. The work of D.P.\ has been supported by the DFG grant LA-905/8-1.

\end{document}